\documentclass{aa}  

\usepackage{graphicx}
\usepackage{txfonts}
\usepackage{hyperref}
\bibpunct{(}{)}{;}{a}{}{,}

\usepackage{amsmath}    
\usepackage{amssymb}    
\usepackage{multicol}        
\usepackage{bm}         
\usepackage{pdflscape}  

\usepackage[T1]{fontenc}

\usepackage{comment}
\usepackage{array}

\newcommand{\xmm}{{\em XMM--Newton}}

\newcommand{\swift}{{\em Swift}}

\newcommand{\Fermi}{{\em Fermi}}

\newcommand{\wise}{{\em WISE}}
\newcommand{\rosat}{{\em ROSAT}}

\newcommand{\aer}[3]{$#1^{+ #2}_{- #3}$}
\newcommand{\aerm}[3]{#1^{+ #2}_{- #3}}
\newcommand{\aerexp}[4]{$#1^{+ #2}_{- #3} \times 10^{#4}$}
\newcommand{\ser}[2]{$#1 \pm #2$}

\newcommand{\serexp}[3]{($#1 \pm #2) \times 10^{#3}$}

\newcommand{\expo}[2]{$ #1 \times 10^{#2}$}
\newcommand{\tento}[1]{$10^{#1}$}
\newcommand{\tentom}[1]{10^{#1}}
\newcommand{\expom}[2]{ #1 \times 10^{#2}}

\newcommand{\aerbm}[3]{$ \mathbf{#1^{+ #2}_{- #3}}$}
\newcommand{\serbm}[2]{$ \mathbf{#1 \pm #2}$}

\newcommand{\pers}{s$^{-1}$}
\newcommand{\chisq}{\chi^{2}}
\newcommand{\rchisq}{\chi^{2}/\textrm{dof}}
\newcommand{\dchi}{\Delta \chi^{2}}
\newcommand{\ddof}{\Delta \textrm{dof}}
\newcommand{\dcash}{\Delta C}
\newcommand{\rcash}{C/\textrm{dof}}

\newcommand{\nh}{N_{\textrm{H}}}

\newcommand{\aaz}{A_{\textrm{N}}}

\newcommand{\subrm}[1]{_{\textrm{#1}}}
\newcommand{\subsc}[1]{_{\textsc{#1}}}
\newcommand{\suprm}[1]{^{\textrm{#1}}}

\newcommand{\xspec}{{\sc xspec}}

\newcommand{\diskbb}{{\sc diskbb}}

\newcommand{\redden}{{\sc redden}}

\newcommand{\zxipcf}{{\sc zxipcf}}
\newcommand{\eplogpar}{{\sc eplogpar}}
\newcommand{\apec}{{\sc apec}}

\newcommand{\ktbb}{kT_{\textrm{BB}}}

\newcommand{\fluxcgs}{ergs~s$^{-1}$~cm$^{-2}$}
\newcommand{\lumcgs}{ergs~s$^{-1}$}

\newcommand{\ovii}{O~{\sc vii}}

\newcommand{\sqcm}{cm$^{-2}$}

\newcommand{\fermi}{\textit{Fermi}~J1544-0639}

\newcommand{\tesc}{t_{\textrm{esc}}}

\begin{document}

\title{An \xmm\ look at the strongly variable radio-weak BL Lac \fermi}
\titlerunning{An \xmm\ look at \fermi}
\authorrunning{F. Ursini et al.}

\author{F. Ursini\inst{1},
        L. Bassani\inst{1},
        F. Panessa\inst{2},
        E. Pian\inst{1},
        G. Bruni\inst{2},
        A. Bazzano\inst{2},
        N. Masetti\inst{1,3},
        K. Sokolovsky\inst{4,5,6}
        and
        P.~Ubertini\inst{1}
}
\institute{
        INAF-Osservatorio di astrofisica e scienza dello spazio di Bologna, Via Piero Gobetti 93/3, 40129 Bologna, Italy.\\
        \email{francesco.ursini@inaf.it}
        \and
        INAF-Istituto di Astrofisica e Planetologia Spaziali, via Fosso del Cavaliere 100, 00133 Roma, Italy.
        \and
        Departamento de Ciencias F\'isicas, Universidad Andr\'es Bello, Fern\'andez Concha 700, Las Condes, Santiago, Chile.
        \and
        Department of Physics and Astronomy, Michigan State University, East Lansing, MI 48824, USA.
        \and
        Astro Space Center of Lebedev Physical Institute, Profsoyuznaya~St.~84/32, 117997~Moscow, Russia.
        \and
        Sternberg Astronomical Institute, Moscow State University, Universitetskii~pr.~13, 119992~Moscow, Russia.
}

   \date{Received ...; accepted ...}


\abstract
{\fermi/ASASSN-17gs/AT2017egv was identified as a gamma-ray/optical transient on May 15, 2017. Subsequent multiwavelength observations suggest that this source may belong to the new class of radio-weak BL Lacs.
}
{We studied the X-ray spectral properties and short-term variability of \fermi\ to constrain the X-ray continuum emission mechanism of this peculiar source.}
{
        We present the analysis of an \xmm\ observation, 56 ks in length, performed on February 21, 2018. 
        }
{The source exhibits  strong X-ray variability, both in flux and spectral shape, on timescales  of $\sim 10$ ks, with a harder-when-brighter behaviour typical of BL Lacs. The X-ray spectrum is nicely described by a variable broken power law, with a break energy of around 2.7 keV consistent with radiative cooling due to Comptonization of broad-line region photons. We find evidence for a `soft excess', nicely described by a blackbody with a temperature of $\sim 0.2$ keV, consistent with being produced by bulk Comptonization along the jet.}
{}


\maketitle

\section{Introduction}\label{sec:intro}
Active galactic nuclei (AGNs) can be divided into two main classes: jetted and non-jetted \citep{padovani2017}. In jetted AGNs, like
blazars, the high-energy emission is dominated by synchrotron and/or synchrotron self-Compton processes from a powerful relativistic jet whose radiation is beamed towards the observer \cite[e.g.][]{blandford&rees,ghisellini1993}. 
According to the currently leading scenario, blazars and radio galaxies belong to the same population and 
their different observational properties are mainly due to the viewing angle
\citep{urry&padovani}.

Blazars are also divided into two subclasses according to their optical spectra: flat-spectrum radio quasars, having strong and broad emission lines, and BL Lacertae objects (BL Lacs) with weak or absent emission lines. However, a continuity is observed in the properties of the spectral energy distribution (SED) of the different subclasses of blazars  known as the blazar sequence \citep{fossati1998,ghisellini1998,donato2001,cavaliere2002}. This indicates that blazars are indeed a single family, but high-power blazars have radiatively efficient accretion discs able to photoionize the broad-line region (BLR), while low-power BL Lacs have inefficient discs 
\citep[][]{ghisellini2008,tavecchio2008}. An alternative, simplified scenario has been proposed by \cite{giommi2012} in which there is no physical link between luminosity and SED shape, so that the observed blazar sequence could be a result of selection effects \citep[but see also][]{ghisellini2017}.

In general, blazars are characterized by strong and rapid variability at all wavelengths \citep[][and references therein]{marscher2016}, and can undergo dramatic outbursts \citep[e.g.][]{pian1998,pian2006,paliya2015}. 
A striking example is SDSS~J124602.54+011318.8, initially reported as an optical transient possibly associated with a gamma-ray burst \citep{vanden2002} and later shown to be a highly variable BL Lac \citep{gal-yam2002}.

On May 15, 2017, \Fermi/LAT detected a transient not associated with any known gamma-ray or X-ray source \citep{ciprini2017}. Named  \fermi, it showed clear detection for two consecutive weeks. A follow-up observation by \swift/XRT detected a new X-ray source, not present in previous catalogues like \textit{ROSAT} \citep{voges1999,boller2016}, at a position corresponding to the optical transient ASASSN-17gs=AT2017egv\footnote{\url{https://wis-tns.weizmann.ac.il/object/2017egv}}, detected on May 25 \citep{ciprini2017}. The host galaxy has been suggested to be 2MASX~J15441967-0649156, with an estimated redshift of 0.171 \citep{z_fermi} and likely of early type \citep{bruni2018}. 
This object is also a weak radio source present in the NRAO \textit{VLA} Sky Survey \citep[NVSS:][]{condon1998} and in the TIFR \textit{GMRT} Sky Survey \citep[TGSS:][]{intema2017}, with a flux density of 46.6 mJy at 1.4 GHz and of 67 mJy at 150 MHz \citep{bruni2018}. 

\cite{bruni2018} conducted a four-month follow-up of \fermi\ in the radio and optical bands, using the Effelsberg 100m single-dish radio telescope and the 2.1m telescope of the Observatorio Astron\'omico Nacional at San Pedro M\'artir. The radio spectrum is flat, consistent with a blazar-like emission, with no evidence of variability up to four months after the burst; moreover, the SED shows a two-hump shape 
with a high-energy peak typical of low-power blazars \citep{bruni2018}.
The radio loudness, estimated as the ratio $R_X$ between the radio and X-ray luminosities \citep{terashima2003}, is $\log R_X = -4.3$, i.e. at the border between the radio-loud and radio-quiet AGN populations.
Finally, the optical spectrum is intermediate between that of a typical BL Lac and a \cite{fanaroff&riley} type I radio galaxy.
\cite{bruni2018} concluded that \fermi\ is a good candidate for the recently discovered class of radio-weak BL Lacs \citep{massaro2017}.
These objects exhibit the multiwavelength properties of BL Lacs, but, contrary to what is commonly observed in jetted AGNs, they lack a strong radio emission. A few radio-weak BL Lac candidates have been reported in the literature, especially from the optical spectroscopy of unidentified gamma-ray sources \citep[][]{masetti2013,paggi2013,ricci2015,marchesini2016}. However, these sources could be confused with white dwarfs or weak emission line quasars \citep{diamond2009,plotkin2010}. As a result, to date only two sources have been robustly claimed to be radio-weak BL Lacs on the basis of a mid-IR colour-based analysis, namely \wise\ J064459.38+603131.7 and \wise\ J141046.00+740511.2 \citep{massaro2017}.
These two sources were found to have mid-IR colours and SEDs consistent with those of BL Lacs, although they lacked counterparts in available radio surveys \citep{massaro2017}.
Among them, one has only a weak radio core while the other is radio-quiet \citep{cao2019}.
Concerning \fermi, its mid-IR colours built with the \wise\ magnitudes in the 3.4 $\mu$m (W1), 4.6 $\mu$m (W2), and 12 $\mu$m (W3) bands are W1--W2=0.55 and W2--W3=2.09, consistent with the  \wise\ blazar strip \citep{massaro2011,massaro2017}.
The very existence of radio-weak BL Lacs challenges the current understanding of jet physics and the unified model of AGNs since they do not form extended radio structures like those of radio-loud/jetted AGNs \citep{massaro2017}.

\begin{figure} 
        \includegraphics[width=\columnwidth]{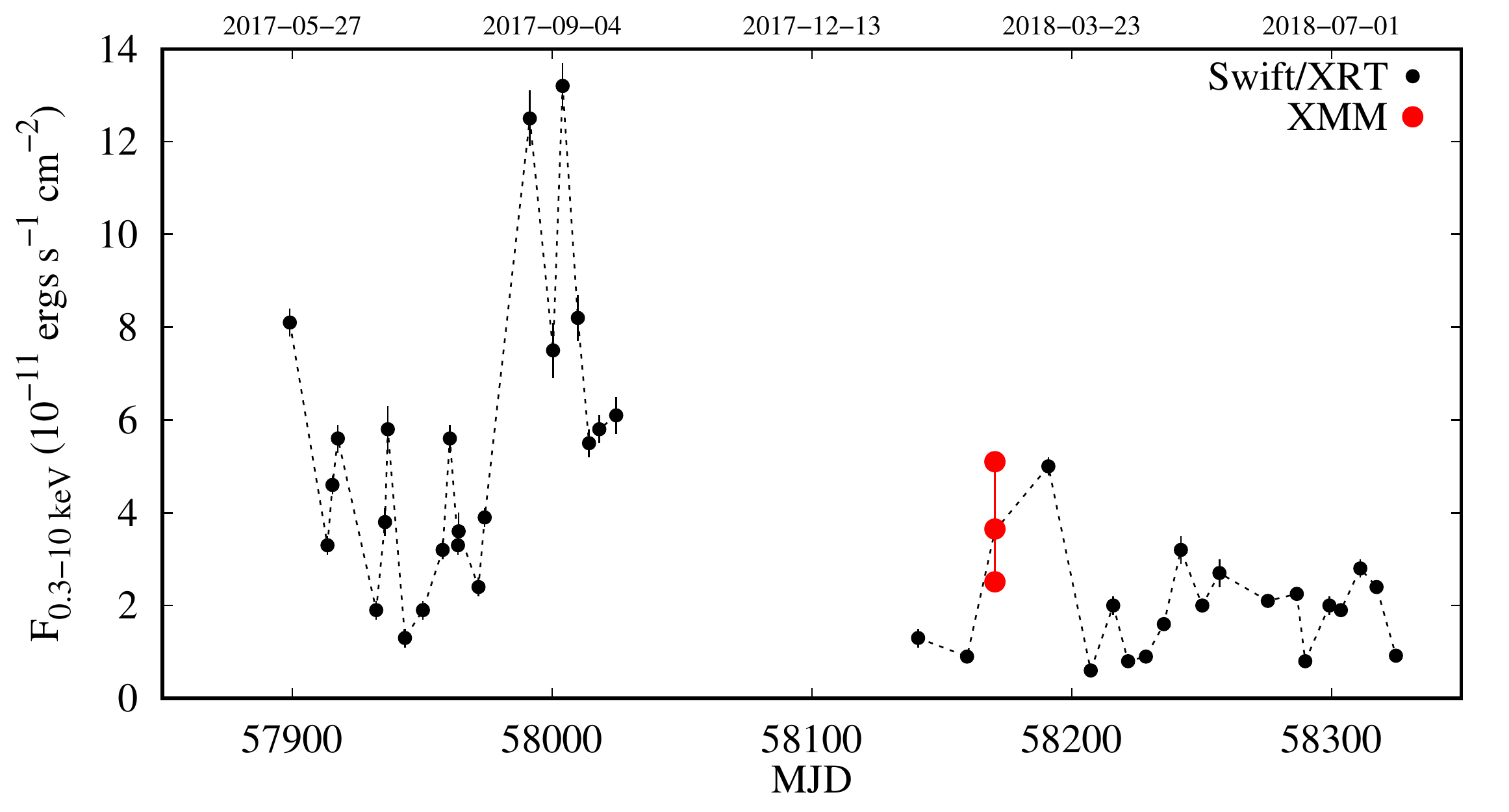}
        \caption{Long-term X-ray light curve of \fermi\ (unabsorbed 0.3--10 keV flux). The black dots denote the \swift/XRT exposures, while the large red  dots denote the \xmm\ observation, separated into three intervals (see Fig. \ref{fig:lc_pn} and Sect. 
        \ref{sec:time-res}). \label{fig:lc_xrt}}
\end{figure}

In X-rays, \fermi\ has been monitored by \swift/XRT starting from May 26, 2017. From the first three-month follow-up, the time-averaged X-ray spectrum was found to be well fitted by a Galactic-absorbed power law with a photon index of \ser{1.86}{0.01} and an unabsorbed 0.3--10 keV flux of \expo{4.7}{-11} \fluxcgs\ \citep{sokolovsky2017}.
However, the source shows a strong flux variability of up to an order of magnitude (see Fig. \ref{fig:lc_xrt}; Sokolovsky et al., in preparation). In February 2018, the XRT monitoring recorded a significant flux decrease 
compared to the average in 2017. 
We thus activated a Target of Opportunity observation with \xmm\ to determine the X-ray spectral properties during 
the low-flux state.
This \xmm\ observation is the main focus of this work.

The paper is organized as follows. In Section \ref{sec:data} we describe the \xmm\ observation and data reduction. In Section \ref{sec:analysis} we discuss the spectral analysis. In Section \ref{sec:discussion} we discuss the results, and in Section \ref{sec:summary} we summarize the conclusions.

\section{Observation and data reduction}\label{sec:data}
\fermi\ was observed by \xmm\ on February 21, 2018 with a net exposure of 56 ks (Obs. Id. 0811213301).
The data were processed using the \xmm\ Science Analysis System (\textsc{sas} v17.0).

\begin{figure} 
        \includegraphics[width=\columnwidth]{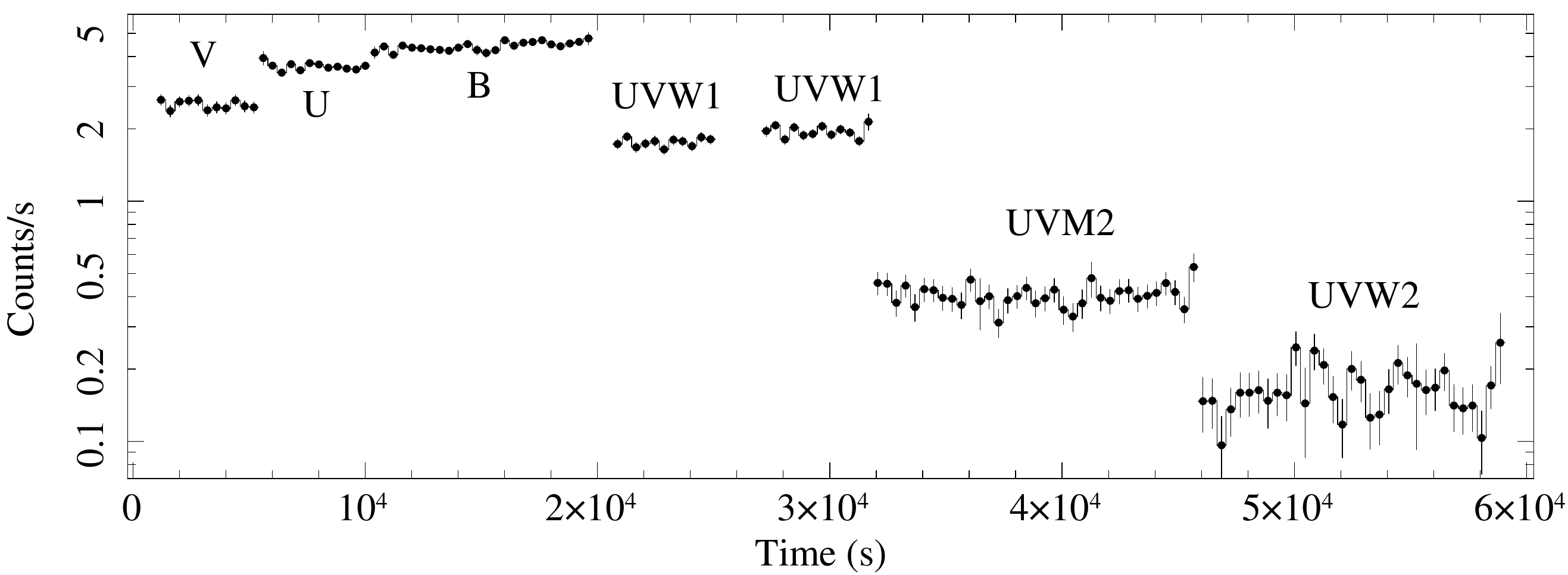}
        \caption{\xmm/OM light curves of \fermi\ for each filter. Bins of 400 s are used. \label{fig:om_lc}}
\end{figure}

The optical monitor \cite[OM;][]{OM} photometric filters were operated in the Science User Defined image/fast mode.
The OM images were taken with all of the six filters (V, B, U, UVW1, UVM2, and UVW2), with a continuous exposure time of 4.4 ks for each image (one image for V and U, two images for B and UVW1, three images for UVM2 and UVW2). 
The OM data were processed with the standard \textsc{sas} pipelines \textsc{omichain} and \textsc{omfchain}, 
and prepared for the spectral analysis using the \textsc{sas} task \textsc{om2pha}. 
We performed the spectral analysis on the stacked exposures for each filter, since no strong and significant temporal variability is observed from the inspection of the OM light curves (plotted in Fig. \ref{fig:om_lc}). The largest variation, namely 10 \%\  in count rate, occurs between the two UVW1 exposures.
 
The EPIC instruments, pn and MOS \cite[][]{pn,MOS}, were operating in the Large Window mode, with the thin filter applied. 
Because of the much higher effective area of the pn detector compared with MOS, throughout the paper we discuss results obtained from pn data. 
However, we checked that the spectral parameters were consistent among MOS and pn, albeit with larger uncertainties in MOS.
Source extraction radii and screening for high-background intervals were determined through an iterative process that maximizes the signal-to-noise ratio \cite[][]{pico2004}. The background was extracted from a circular region with a radius of 50 arcsec, while the source extraction radii were allowed to be in the range 20--40 arcsec; the best extraction radius was found to be 40 arcsec. The light curves were corrected and background-subtracted using the \textsc{sas} task \textsc{epiclcorr}. The EPIC-pn spectra were grouped such that each spectral bin contained at least 30 counts, and did not oversample the instrumental energy resolution by a factor greater than 3. 
We obtained poor fits near 1.8 and 2.2 keV, namely the energies of the instrumental silicon and gold edges. For this reason, we excluded the 1.75--2.25 keV range from the fits. 
For the spectral analysis, we used the full pn band 0.3--10 keV.

\begin{figure} 
        \includegraphics[width=\columnwidth]{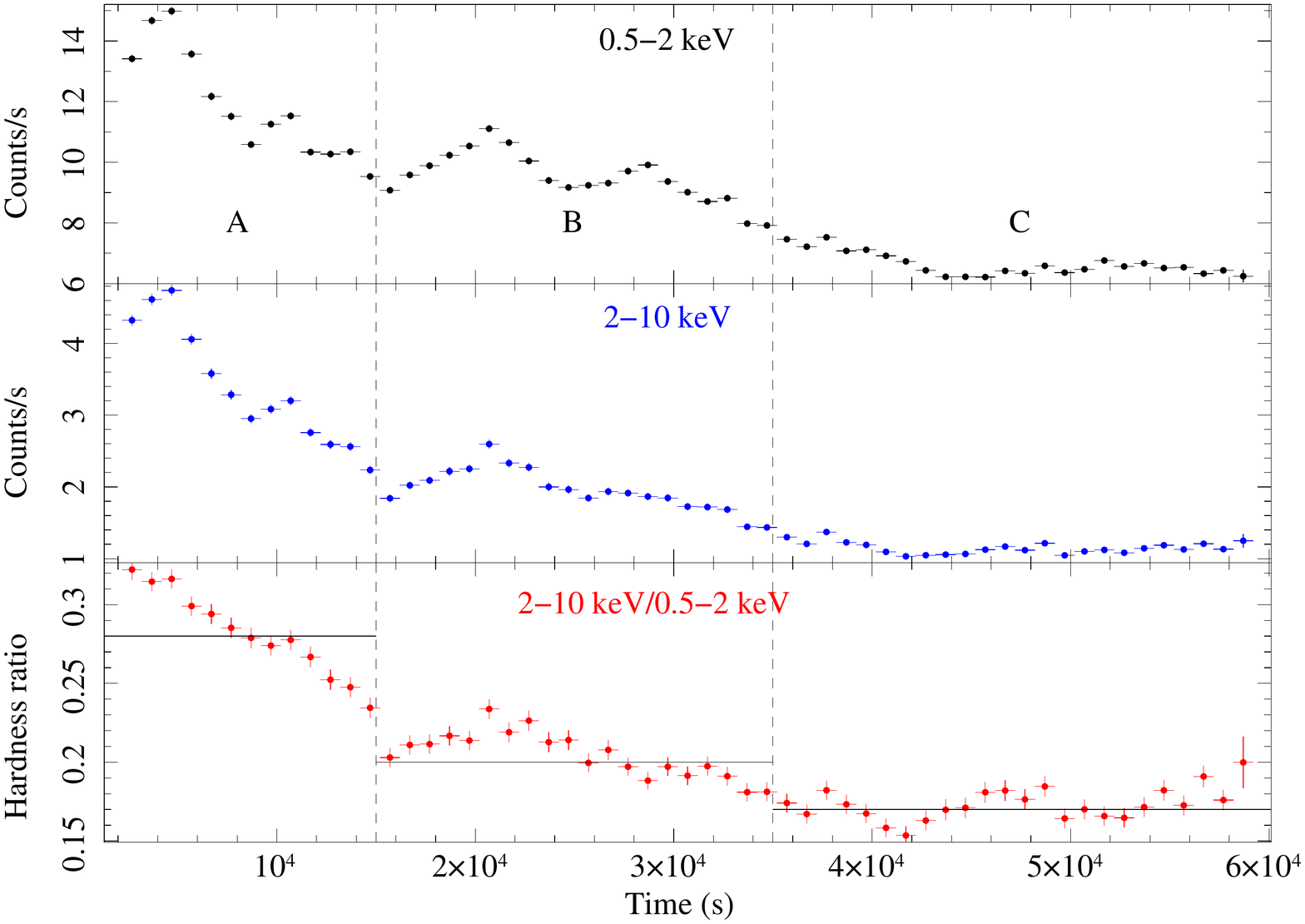}
        \caption{\xmm/pn light curve of \fermi. Bins of 1 ks are used. Top panel: Count rate light curve in the 0.5--2 keV range. Middle panel: Count rate light curve in the 2--10 keV range. Bottom panel: Hardness ratio (2--10/0.5--2 keV) light curve. The vertical dashed lines separate the three intervals (A, B, C) defined for the spectral analysis. The solid horizontal lines correspond to the average hardness ratio for each interval. \label{fig:lc_pn}}
\end{figure}

\begin{figure} 
        \includegraphics[width=\columnwidth]{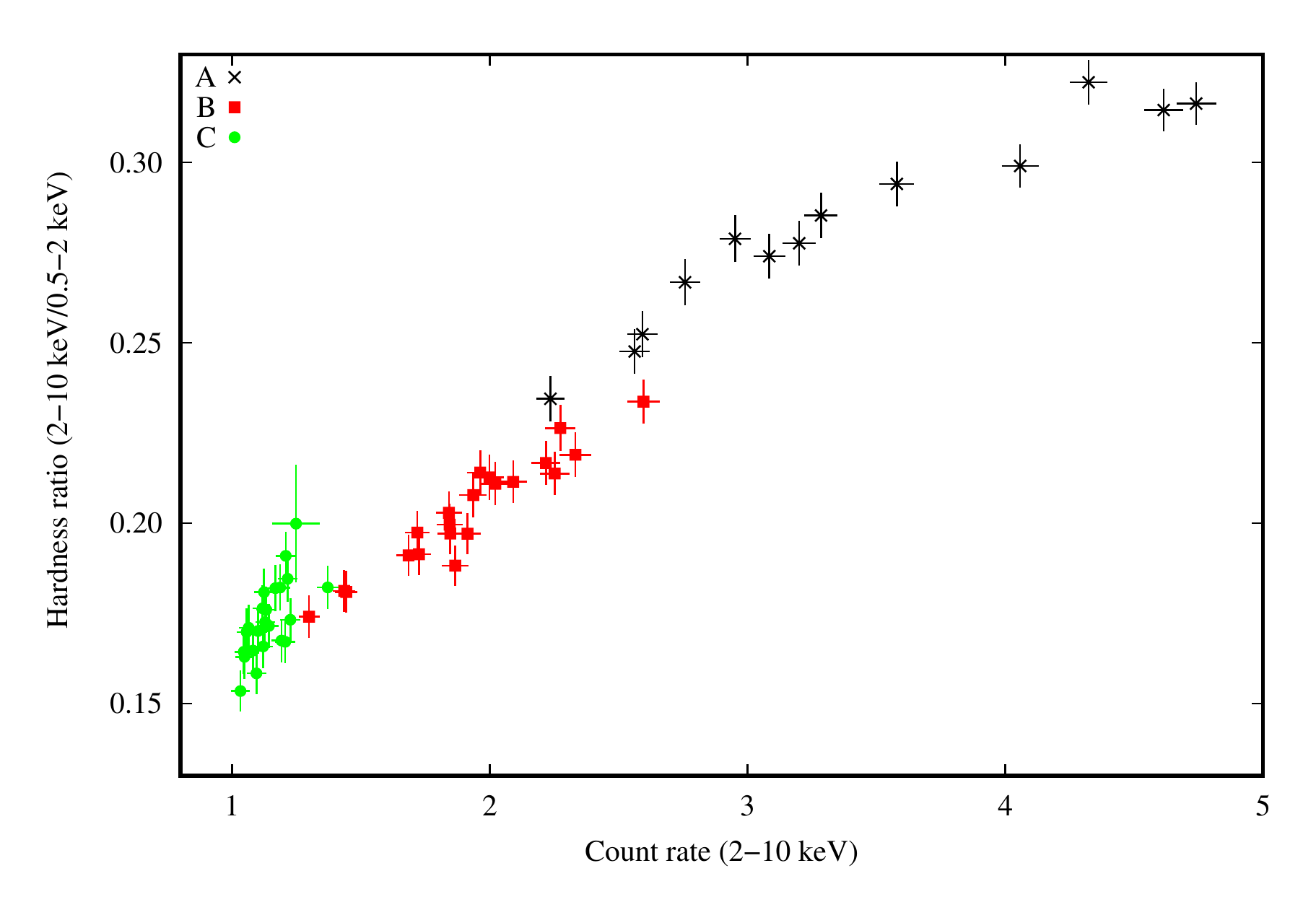}
        \caption{ Hardness ratio plotted against the 2--10 keV count rate for the three intervals A (black), B (red), and C (green). Time bins of 1 ks are used.  \label{fig:HI}}
\end{figure}

In Fig. \ref{fig:lc_pn} we show the \xmm/pn light curves in the 0.5--2 keV and 2--10 keV energy ranges with the hardness ratio,
background subtracted and corrected using the standard \textsc{sas} tool \textsc{epiclccorr}. We use time bins of 1 ks to have uncertainties of a few percentage points in each bin. 
The source is strongly variable, both in flux (up to a factor of 4 in the 2--10 keV count rate) and in spectral shape,  with a clear `harder-when-brighter' trend. We plot the hardness-intensity diagram, namely the hardness ratio versus the 2--10 keV count rate, in Fig. \ref{fig:HI}.
We calculated the fractional rms variability
amplitude \citep[$F_{\textrm{var}}$; e.g.][]{vaughan2003}
using the tool \textsc{lcstats} in \textsc{xronos}. For the 0.5--2 keV light curve and the 2--10 keV light curve we found an $F_{\textrm{var}}$ of \ser{0.26}{0.02} and \ser{0.49}{0.05}, respectively.
To perform a time-resolved spectral analysis, we split the total exposure into three intervals, labelled  A (15 ks long), B (20 ks), and C (21 ks), on the basis of the hardness ratio. The mean 2--10 keV/0.5--2 keV hardness ratio for the intervals A, B, and C is 0.28, 0.20, and 0.17, respectively. The first interval (int. A) thus corresponds to the harder/brighter state, while the third  (int. C) corresponds to the softer/fainter state. 
This choice allows us to explore different spectral states with comparable statistics. We list the observed number of EPIC-pn counts for each interval in Table \ref{tab:counts}, together with the net exposure.

\begin{table}
        \begin{center}
                \caption{ \label{tab:counts} Net exposure and number of pn counts in the 0.3--10 keV band for the different intervals.}
                \begin{tabular}{ c c c  } 
                        \hline 
                        interval & net exp. (ks) & counts \\ 
                        \hline
                        A & 14& 188 105 \\
                        B & 18& 200 527 \\
                        C & 19& 145 472 \\              
                        \hline          
                \end{tabular}
        \end{center}
\end{table}

\section{Spectral analysis}\label{sec:analysis}
We performed the spectral analysis with the \xspec\ 12.10 package \cite[][]{arnaud1996}. 
The binned pn spectra and the OM photometric data were analysed using the $\chisq$ minimization technique. All errors are quoted at the 90 \%\ confidence level ($\dchi = 2.71$) for one interesting parameter. 
In the fits we always included neutral absorption ({\sc phabs} model in {\sc xspec}) from Galactic hydrogen with column density $\nh = 8.7 \times 10^{20}$ \sqcm\ \cite[][]{kalberla2005}. 
We assumed the element abundances of \cite{angr} and the photoelectric absorption cross sections of \cite{vern}.

\begin{figure} 
        \includegraphics[width=\columnwidth]{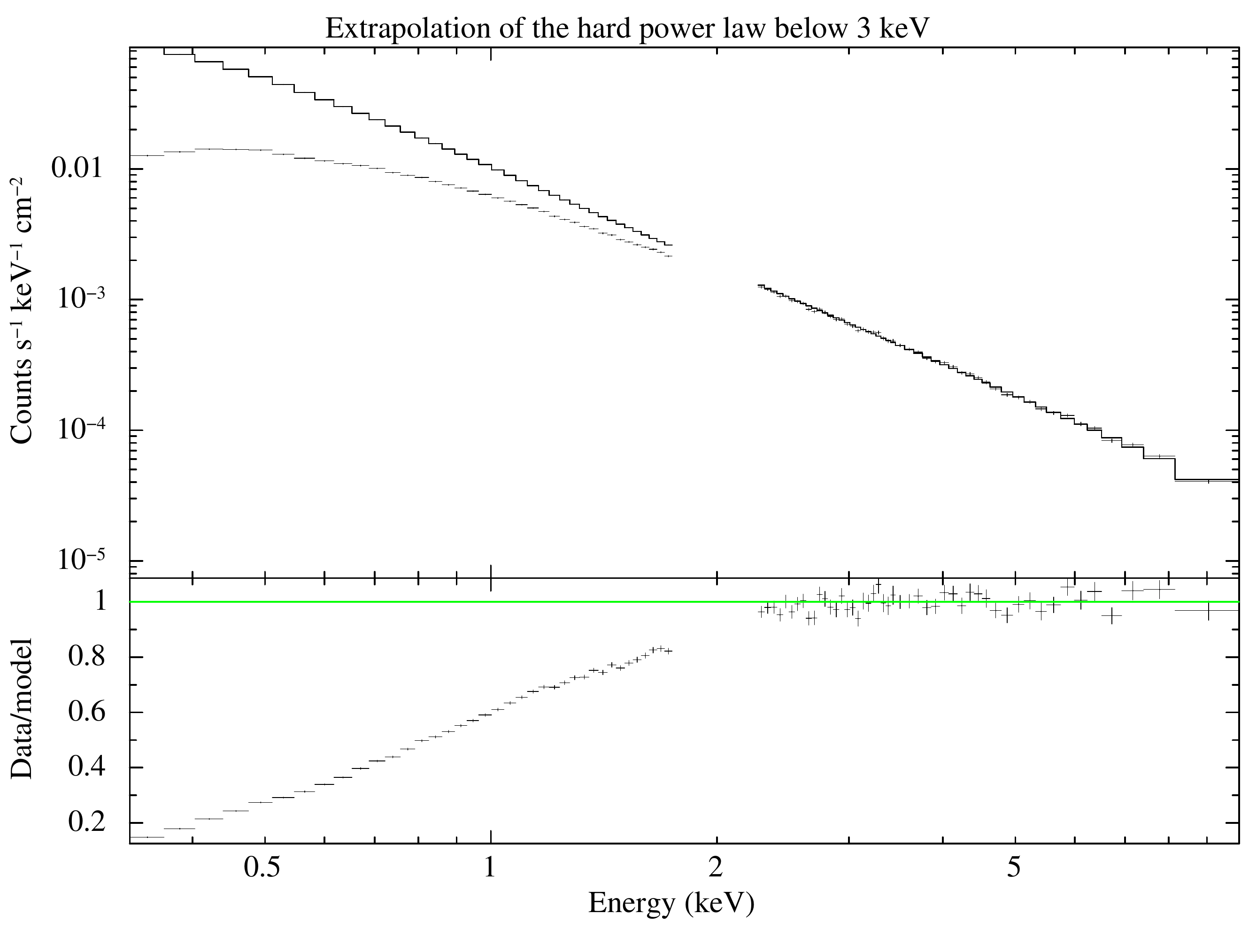}
        \caption{Upper panel: Time-averaged \xmm/pn spectrum of \fermi. The power law that best fits the data in the 3--10 keV band is shown as a continuous curve. Lower panel: Ratio of the observed spectrum (0.3--10 keV) to best-fit 3--10 keV power law. The data were binned for plotting purposes. \label{fig:extrap}}
\end{figure}

\subsection{Time-averaged spectrum}\label{sec:pn_average}
First, we fitted the average pn spectrum integrated over the full exposure. As a first step, we fitted the pn spectrum above 3 keV with a simple power law. Extrapolating this best-fitting power law to lower energies, we note a clear spectral break with a significant flattening below 2 keV (Fig. \ref{fig:extrap}).

\begin{figure} 
        \includegraphics[width=\columnwidth]{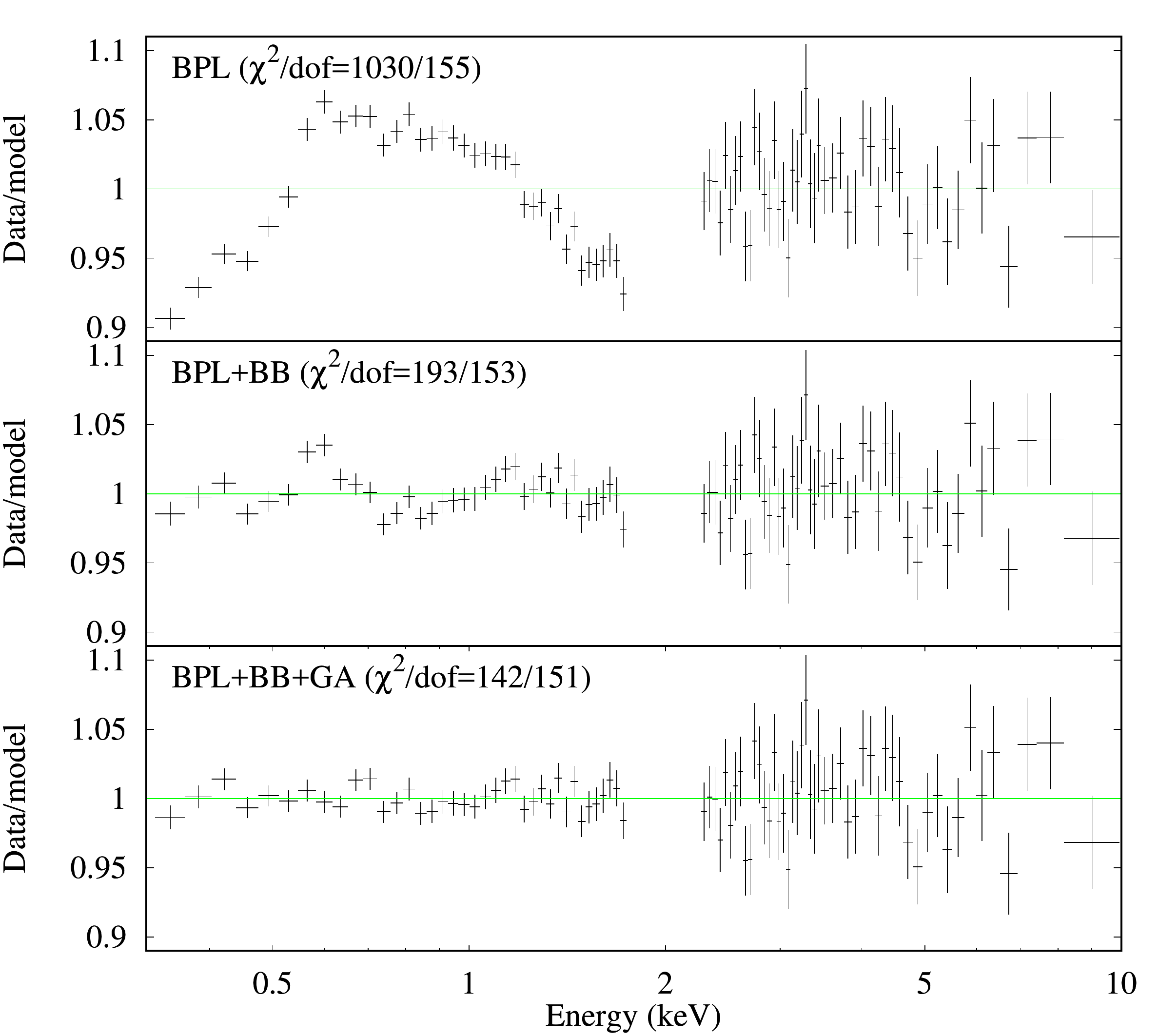}
        \caption{Residuals of the fits of the time-averaged pn spectrum in the 0.3--10 keV band with the BPL model (Sect. \ref{sec:pn_average}). Top panel: Simple broken power law. Middle panel: Broken power law plus blackbody. Bottom panel: Broken power law plus blackbody and a Gaussian line at 0.7 keV rest-frame. The data were binned for plotting purposes. \label{fig:ratio_average}}
\end{figure}

Then we fitted the spectrum with a broken power law (BPL), characterized by the break energy $E_b$ and by the photon indices $\Gamma_1$ below the break and $\Gamma_2$ above the break. We found a poor fit, with $\rchisq=1030/155$ and significant residuals with a bumpy shape below 2 keV (Fig. \ref{fig:ratio_average}, upper panel).
To account for this `soft excess', 
we tried to include a further absorption component with \textsc{zphabs} in \xspec. We first assumed a zero redshift, corresponding to an excess of Galactic absorption,  finding  a poor fit ($\rchisq = 492/154 $). Next, we assumed the redshift of the source, finding again a poor fit ($\rchisq = 360/154 $). Then, we removed the absorber and
we included a  blackbody with a free temperature $\ktbb$. We found a much better fit ($\rchisq=193/153$, i.e. $\dchi/\ddof=-837/-2$) and $\ktbb \simeq 0.2$ keV, but still with significant positive residuals around 0.6 keV (Fig. \ref{fig:ratio_average}, middle panel). The best-fitting parameters are listed in Table \ref{tab:fits_pn}.

To account for the residuals at 0.6 keV, we first included an ionized absorber (\zxipcf\ in \xspec) with free covering fraction, column density, and ionization parameter. We found a better fit ($\rchisq=157/150$, i.e. $\dchi/\ddof=-46/-3$), 
whose parameters are listed in Table \ref{tab:fits_pn}.
However, we still found some positive residuals at 0.6 keV. We thus included a narrow Gaussian emission line, finding a better fit; in this case, the absorption component is not needed and its covering fraction is pegged at zero. We thus removed the ionized absorber, finally obtaining $\rchisq=142/151$, i.e. $\dchi/\ddof=-15/+1$ (Fig. \ref{fig:ratio_average}, lower panel; the probability of chance improvement is  \expo{9}{-11} from an $F$-test). 
The best-fitting parameters are given in Table \ref{tab:fits_pn}.
The line is found to have a rest-frame energy of \aer{0.70}{0.02}{0.01} keV and, if physical, might be interpreted as a blend of the two $2p{-}3s$ lines of \ion{Fe}{xvii} at $17.05{-}17.10$ \AA. 
In Sect. \ref{sec:time-res} we  show that neither the soft excess nor the emission line are artefacts of variability. 

Furthermore, we checked whether the line detection is affected by the assumed elemental abundances. 
For this test, we switched to the abundances of \cite{asplund2009}. We obtained a slightly worse fit with $\rchisq=155/151$ and residuals at 0.3--0.4 keV suggestive of a slight excess of absorption. We thus added a further \textsc{phabs} component, finding a good fit and no need for the Gaussian emission line. Removing the line, we obtained $\rchisq=144/152$ and a column density of \serexp{2.1}{0.8}{20} \sqcm\ in excess of the value of \cite{kalberla2005}. The other parameters of the fit are not strongly altered by the different abundances. Analogous results were found assuming the abundances of \cite{lodders2003}. Therefore, we cannot exclude that the residuals at 0.6 keV (observer-frame) are due to an imperfect modelling of Galactic absorption.  However, in the subsequent analysis we decide to assume the \cite{angr} abundances and modelled the residuals with a phenomenological Gaussian emission line, which is more consistent with the RGS data (see Appendix A).
Finally, to test whether the putative emission line is consistent with an origin from the hot interstellar medium of the host galaxy, we replaced it with the thermal plasma model \apec, v3.0.9 \citep{apec}. We obtained a good fit ($\rchisq=149/151$).
In general, the element abundances are consistent with being solar. However, some positive residuals near 0.4 keV (rest-frame) could indicate an enhanced emission from \ion{N}{vi}. Leaving the nitrogen abundance (relative to solar) $\aaz$ free, we obtain $\rchisq=141/150$ ($\dchi/\ddof=-8/-1$) and $\aaz=\aerm{17}{6}{7}$ times solar. The other parameters are given in Table \ref{tab:fits_pn}. We note that the \apec\ component alone is not able to reproduce the soft excess; indeed, removing the blackbody results in a poor fit ($\rchisq=355/152$). 

The intrinsic (i.e. unabsorbed) flux in the 0.3--10 keV band is \serexp{3.61}{0.01}{-11} \fluxcgs, not far from the average of \expo{4.7}{-11} \fluxcgs\ recorded by \swift/XRT in summer 2017 \citep[][see also Fig. \ref{fig:lc_xrt}]{sokolovsky2017}.

To try to further characterize the curvature of the X-ray spectrum, we fitted the data with a log-parabolic spectral law.
We obtained similar results concerning the presence of the soft excess, although the final fit is statistically worse than the BPL best fit. This analysis is discussed in detail in Appendix B.

\begin{table*}
        \begin{center}
                \caption{ \label{tab:fits_pn} Parameters of the best-fitting spectral models for the time-averaged pn spectrum (Sect. \ref{sec:pn_average}). First column: \textsc{bknpow+bbody}, second column: \textsc{zxipcf*(bknpow+bbody)}, third column: \textsc{bknpow+bbody+zgauss}, fourth column: \textsc{bknpow+bbody+apec}. The energy of the Gaussian line is rest-frame. The column in boldface corresponds to the statistically preferred model.} 
                \begin{tabular}{ l c c >{\bf}c c } 
                        \hline  
                         & BPL+BB & ZXIPCF & BPL+BB+GA & BPL+BB+APEC \\
                        \hline
                        $\Gamma_1$ &\ser{1.88}{0.03}&\ser{1.82}{0.04}&\serbm{1.93}{0.03}&\aer{1.84}{0.07}{0.09} \\
                        $E_{b}$ &\ser{2.55}{0.09}&\ser{2.49}{0.09}&\serbm{2.69}{0.12}&\ser{2.80}{0.14} \\
                        $\Gamma_2$ &\ser{2.52}{0.02}&\ser{2.53}{0.02}&\serbm{2.52}{0.02}&\ser{2.51}{0.02} \\
                        $F\subrm{1keV}$ ($\mu$Jy) &\ser{5.77}{0.09}&\ser{5.66}{0.14}&\serbm{5.84}{0.08}&\ser{5.77}{0.10} \\
                        \noalign{\medskip}
                        $\ktbb$ (keV)&\ser{0.203}{0.006}&\ser{0.206}{0.006}&\serbm{0.220}{0.009}&\ser{0.250}{0.011}\\
                        $N_{\textrm{BB}}$ (\tento{-5})&\ser{7.4}{0.6}&\ser{9.1}{0.8}&\serbm{6.5}{0.5}&\ser{7.6}{1.2}\\
                        \noalign{\medskip}
                        $E_{\textrm{GA}}$ (keV)&-&-&\aerbm{0.70}{0.02}{0.01}&- \\      
                        $N_{\textrm{GA}}$ (\tento{-4}) &-&-&\serbm{1.6}{0.4}&- \\      
                        \noalign{\medskip}
                        $\nh$ (\tento{22} \sqcm) &-&\aer{4}{8}{2}&-&- \\        
                        $\log \xi$ &-&\aer{3.3}{0.1}{0.2}&-&- \\
                        $C_F$ &-&\ser{0.4}{0.3}&-&-\\   
                        \noalign{\medskip}
                        $kT\subsc{apec}$ (keV)&-&-&-&\ser{0.27}{0.01}\\
                        $N\subsc{apec}$ (\tento{-4})&-&-&-&\ser{7}{2}\\ 
                        $\aaz$ (solar) &&&&\aer{17}{6}{7}\\
                        \noalign{\medskip}
                        $\rchisq$ &193/153&157/150&142/151&141/150 \\                                           
                        \hline          
                \end{tabular}
        \end{center}
\end{table*}

Next, we fitted the time-averaged pn spectrum jointly with the optical/UV data from the OM. 
In these fits we included interstellar extinction (\redden\ model in \xspec) with $E(B-V)=0.15$ \cite[]{schlegel1998}. We also included the host galaxy contribution, employing a template spectrum of an elliptical galaxy \citep{kinney1996}
and leaving its normalization free to vary. 

\begin{figure} 
        \includegraphics[width=\columnwidth]{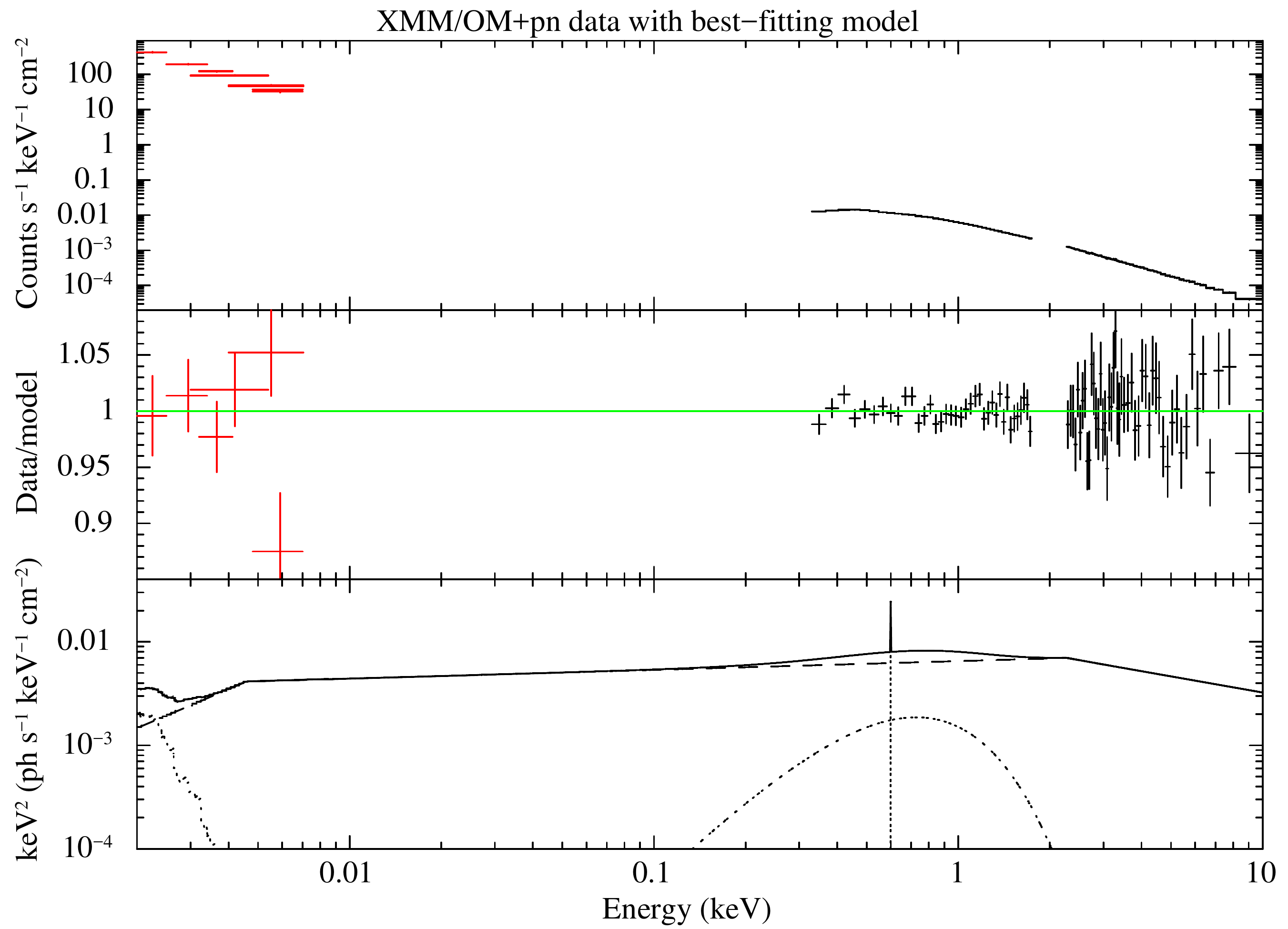}
        \caption{Time-averaged \xmm\ spectrum with best-fitting 2-BPL model (Sect. \ref{sec:pn_average}). Upper panel: OM (red) and pn (black) data with folded model. Middle panel: Data-to-model ratio. Lower panel: Best-fitting model $E^2f(E)$, without Galactic absorption and reddening, including the double-broken power law (dashed line) plus the blackbody, the Gaussian line, and the host galaxy (dotted lines). The data were binned for plotting purposes. \label{fig:fit_broad}}
\end{figure}

We started from the BPL best fit of the pn spectrum including the blackbody and Gaussian line. This model yields a poor fit in the OM band ($\rchisq=221/156$). Then we tested for the presence of a further emission component in the optical/UV band, which we modelled with a multicolour disc blackbody \cite[\diskbb\ in \xspec:][]{mitsuda1984,makishima1986}.
We found an improved but still unacceptable fit with $\rchisq=189/154$. This is likely due to the fact that the primary X-ray power law cannot extend down to the optical/UV band. Therefore, we tested a different model allowing for a second spectral break at low energies.
We thus replaced the broken power law with a double-broken power law (2-BPL). The 2-BPL is characterized by three different photon indices ($\Gamma_0$, $\Gamma_1$, and $\Gamma_2$) and by two different break energies ($E_{b,1}$ and $E_{b,2}$). First, we assumed a negligible optical/UV contribution from the X-ray power law by freezing the low-energy break $E_{b,1}$ at 0.1 keV. We obtained a good fit with $\rchisq=158/153$ ($\dchi/\ddof=-31/-1$). 
Then we left the low-energy break of the 2-BPL free. We obtained a further improvement and no need for the \diskbb\ component, whose normalization became consistent with zero. We thus removed the \diskbb\ and obtained $\rchisq=151/154$ ($\dchi/\ddof=-7/+1$), with a low-energy break at $\sim 5$ eV.
In Fig. \ref{fig:fit_broad} we show the data with residuals and best-fitting model, while the best-fitting parameters are given in Table \ref{tab:fit_broad}.

\begin{table*}
        \begin{center}
                \caption{ \label{tab:fit_broad} 
                        Parameters of the best-fitting spectral models for the time-averaged OM+pn spectrum (Sect. \ref{sec:pn_average} and Fig. \ref{fig:fit_broad}). Column 1: \textsc{bknpow+bbody+zgauss};  Column 2: \textsc{diskbb+bknpow+bbody+zgauss}; Column 3: \textsc{diskbb+bkn2pow+bbody+zgauss};  Column 4: \textsc{bkn2pow+bbody+zgauss}. The energy of the Gaussian line is rest-frame. (f) denotes a frozen parameter. Column 4 (in bold) is the statistically preferred model.}
                \begin{tabular}{ l c c c >{\bf}c } 
                        \hline  
                        &BPL & BPL+DISK & 2-BPL+DISK & 2-BPL \\
                        \hline
                        $kT\subsc{diskbb}$ (eV) &-&\ser{42}{3}&\ser{3.8}{0.6}&- \\
                        $N\subsc{diskbb}$ &-&\serexp{1.6}{0.5}{6}&\aerexp{4}{3}{1}{9}&-\\
                        \noalign{\medskip}
                        $\Gamma_0$ &-&-&1(f)&\aerbm{0.8}{0.3}{0.5} \\
                        $E_{b,1}$ (eV)&-&-& 100(f)& \aerbm{5.3}{0.5}{0.4}\\
                        $\Gamma_1$ &\ser{1.884}{0.005}&\ser{1.82}{0.02}& \ser{1.93}{0.03} & \serbm{1.92}{0.01}\\ 
                        $E_{b,2}$ (keV)&\ser{2.58}{0.07}&\ser{2.51}{0.07}& \ser{2.69}{0.12} & \serbm{2.66}{0.08} \\                                
                        $\Gamma_2$ &\ser{2.52}{0.02}&\ser{2.52}{0.02}&\ser{2.52}{0.02} & \serbm{2.52}{0.02} \\         
                        $F\subrm{1keV}$ ($\mu$Jy) &\ser{5.76}{0.06}&\ser{5.5}{0.1}&\ser{5.84}{0.08}&\serbm{5.83}{0.06} \\
                        \noalign{\medskip}
                        $\ktbb$ (keV)&\ser{0.209}{0.004}&\ser{0.205}{0.005}& \ser{0.220}{0.007} &\serbm{0.217}{0.005} \\     
                        $N_{\textrm{BB}}$ (\tento{-5})&\ser{7.1}{0.03}&\ser{8.3}{0.5}&\ser{6.5}{0.4}  & \serbm{6.6}{0.3} \\  
                        \noalign{\medskip}
                        $E_{\textrm{GA}}$ (keV)&\aer{0.70}{0.02}{0.01}&\aer{0.69}{0.02}{0.01}&\aer{0.70}{0.02}{0.01} &\aerbm{0.70}{0.02}{0.01} \\    
                        $N_{\textrm{GA}}$ (\tento{-4})&\ser{1.5}{0.4}&\ser{1.6}{0.4}& \ser{1.6}{0.4} &\serbm{1.6}{0.4} \\                             
                        \noalign{\medskip}
                        $\rchisq$&221/156&189/154& 158/153 &151/154 \\                                                                  
                        \hline          
                \end{tabular}
        \end{center}
\end{table*}

Extrapolating the best-fitting model to the boundaries of the 0.001--10 keV band, we find an intrinsic luminosity of \serexp{6.22}{0.08}{45} \fluxcgs.


\subsection{Time-resolved spectral analysis of pn data}\label{sec:time-res}
Having obtained a characterization of the time-integrated spectrum, we moved to the analysis of the pn spectra of the three intervals A, B, and C separately.  
Since our goal was to study the spectral variability in the X-ray band, we did not include the OM data for simplicity.

\begin{figure} 
        \includegraphics[width=\columnwidth]{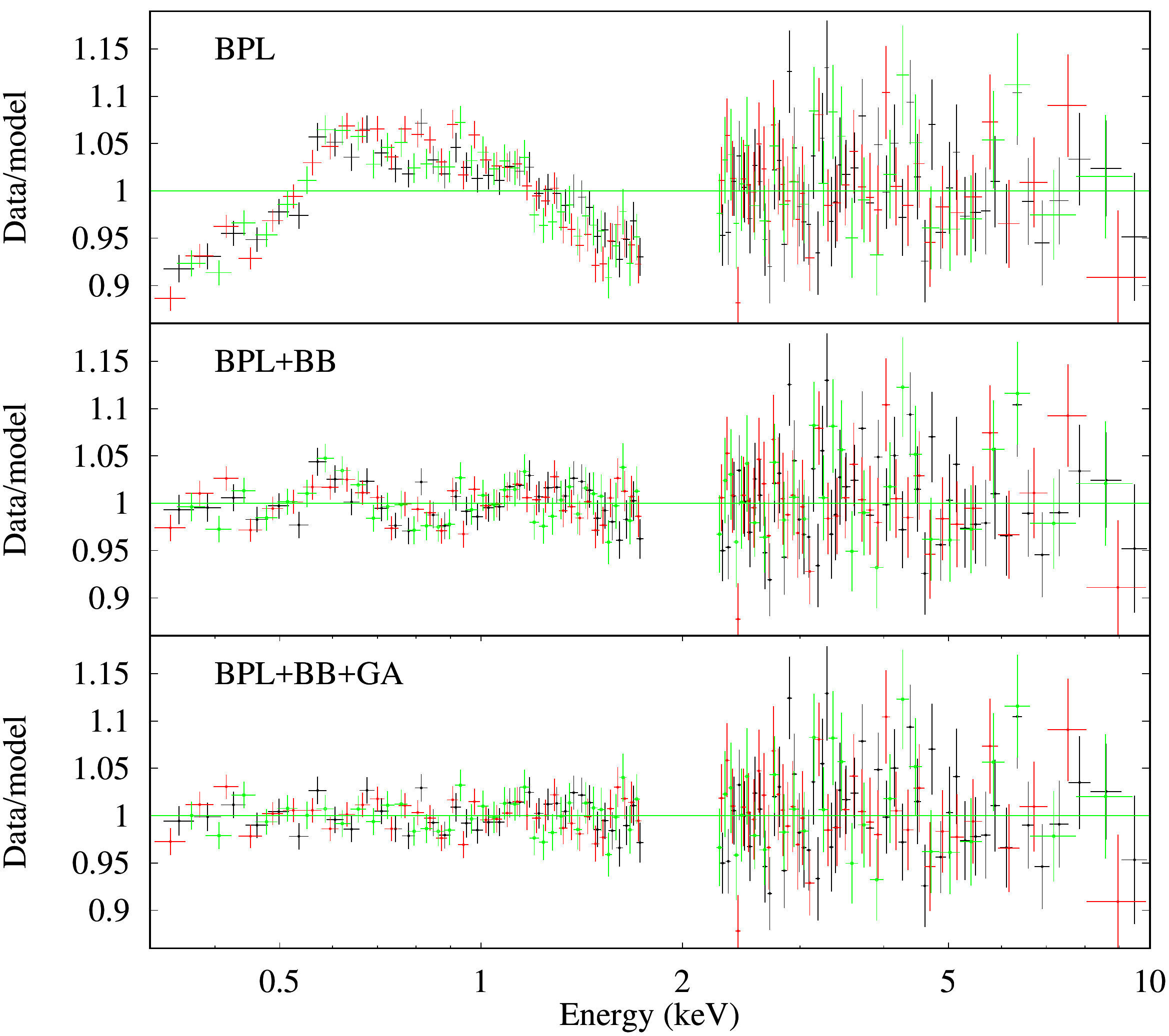}
        \caption{Residuals of the fits of the three pn spectra in the 0.3--10 keV band with the BPL model (Sect. \ref{sec:time-res}), for interval A (black), B (red), and C (green). Upper panel: Simple broken power law. Middle panel: Broken power law plus blackbody. Lower panel: Broken power law plus blackbody and Gaussian line. The data were binned for plotting purposes. \label{fig:res}}
\end{figure}

\begin{figure} 
        \includegraphics[width=\columnwidth]{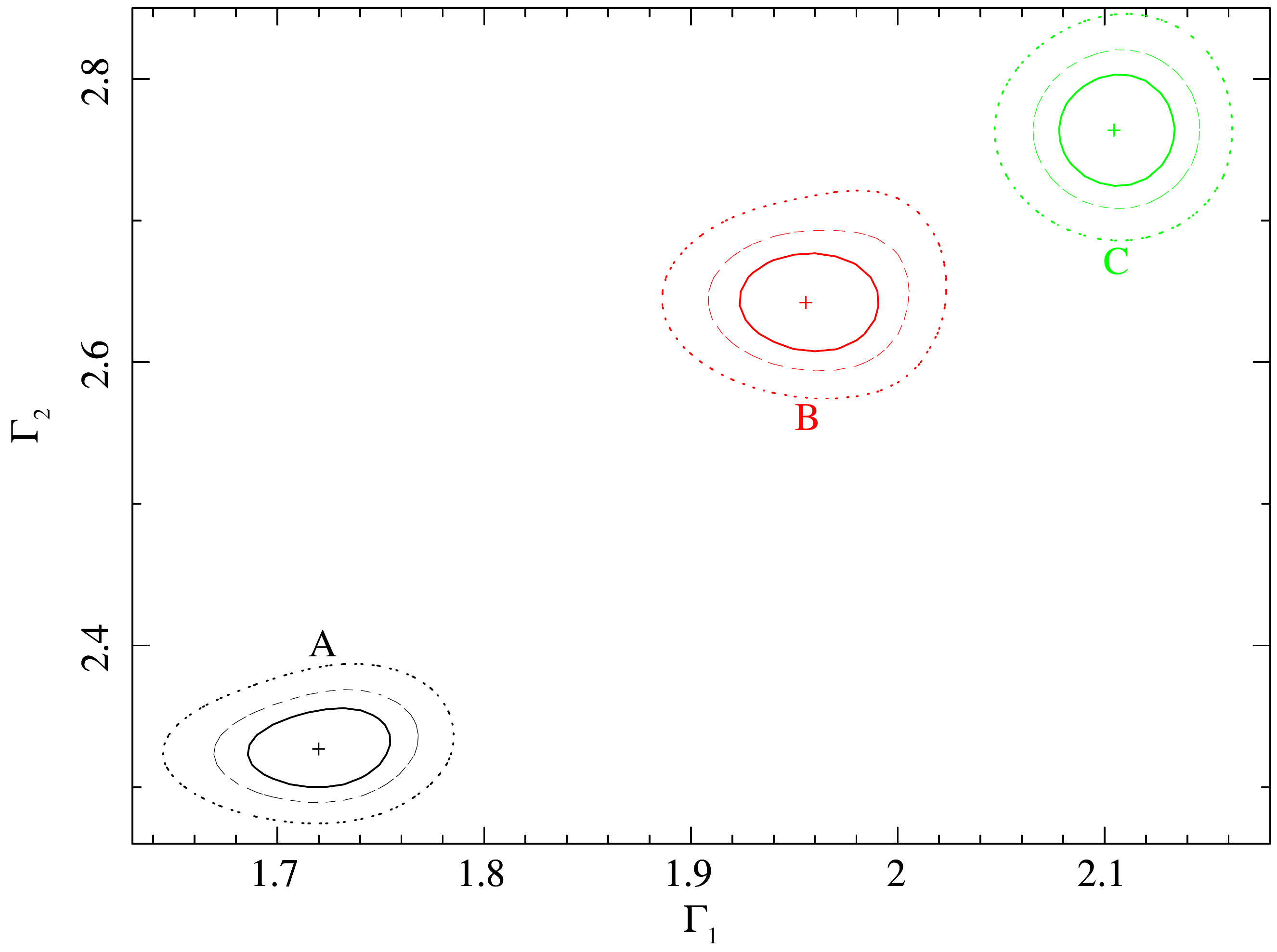}
        \caption{Contour plots of the hard photon index $\Gamma_2$ vs. the soft photon index $\Gamma_1$ of the BPL model for interval A (black), B (red), and C (green). Solid, dashed, and dotted lines correspond to 68 \%, 90 \% and 99 \%\ confidence levels, respectively.  \label{fig:cont}}
\end{figure}

As in Sect. \ref{sec:pn_average}, we started by fitting the three pn spectra with the BPL model. We left the photon indices ($\Gamma_1$ and $\Gamma_2$), the break energy ($E_b$), and the normalization of the BPL free to vary among the different intervals. 
We found a poor fit, with $\rchisq=1347/422$ and significant residuals with a bumpy shape below 2 keV (see Fig. \ref{fig:res}). To account for these residuals, we included a blackbody, leaving its normalization free to vary among the three spectra; the temperature $\ktbb$ was free but tied among the three intervals, with no detectable improvement by allowing it to change. We found a much better fit ($\rchisq=490/418$, i.e. $\dchi/\ddof=-857/-4$), still with significant residuals around 0.6 keV (Fig. \ref{fig:res}). A further improvement was found by adding a constant, narrow Gaussian emission line ($\rchisq=441/416$, i.e. $\dchi/\ddof=-49/-2$; the probability of chance improvement is \expo{3}{-10} from an $F$-test). 
The best-fitting parameters are listed in Table \ref{tab:fits1}, while in Fig. \ref{fig:cont} we show the contour plots of the two photon indices for the three intervals. The difference $\Delta \Gamma = \Gamma_2 - \Gamma_1$ is around 0.6--0.7.
We also tested an absorption scenario to explain the residuals in the soft X-ray band. We thus removed the Gaussian line and included an ionized absorber (\zxipcf\ in \xspec). The covering fraction, column density, and ionization parameter were tied among the intervals. We found a worse fit with $\rchisq=454/415$, i.e. $\dchi=+13/-1$, and no statistical improvement by allowing the absorber parameters to change among the intervals. Finally, we included an \apec\ component instead of the single line, finding $\rchisq=444/415$ and a nitrogen abundance $\aaz=\aerm{16}{8}{10}$ relative to solar.


\begin{table}
        \begin{center}
                \caption{ \label{tab:fits1} Best-fitting parameters for the time-resolved pn spectra (BPL model: \textsc{bknpow+bbody+zgauss}, see Sect. \ref{sec:time-res}). The energy of the Gaussian line is rest-frame. (t) denotes a parameter tied among the different intervals.}
                \begin{tabular}{ l c c c  } 
                        \hline  
                        & A & B & C \\
                        $\Gamma_1$ &\ser{1.72}{0.04}&\aer{1.96}{0.03}{0.04}&\ser{2.11}{0.03} \\      
                        $E_b$ (keV) &\aer{2.6}{0.2}{0.1}&\aer{2.7}{0.2}{0.1}&\aer{2.6}{0.1}{0.2} \\                              
                        $\Gamma_2$ &\ser{2.33}{0.03}&\ser{2.64}{0.04}&\ser{2.76}{0.04} \\              
                        $F\subrm{1keV}$ ($\mu$Jy) &\ser{7.4}{0.2}&\ser{6.0}{0.1}&\ser{4.5}{0.1} \\
                        \noalign{\medskip}
                        $\ktbb$ (keV)&\aer{0.218}{0.010}{0.008}&(t)&(t) \\      
                        $N_{\textrm{BB}}$ (\tento{-5}) &\ser{7.7}{1.0}&\ser{7.8}{0.08}&\ser{4.7}{0.6} \\      
                        \noalign{\medskip}
                        $E_{\textrm{GA}}$ (keV)&\aer{0.72}{0.02}{0.01}&(t)&(t) \\      
                        $N_{\textrm{GA}}$ (\tento{-4}) &\ser{1.5}{0.4}&(t)&(t) \\                              
                        \noalign{\medskip}
                        $F_{\textrm{0.3--2 keV}}$ (\tento{-11} &\ser{2.87}{0.02}&\ser{2.44}{0.01}&\ser{1.80}{0.01} \\
                        \fluxcgs)&&&\\
                        $F_{\textrm{2--10 keV}}$ (\tento{-11} &\ser{2.20}{0.02}&\ser{1.20}{0.02}&\ser{0.71}{0.01} \\
                        \fluxcgs)&&&                            
                        \\ 
                        \noalign{\medskip}
                        $\rchisq$&441/416&&\\                                                                   
                        \hline          
                \end{tabular}
        \end{center}
\end{table}

\section{Discussion}\label{sec:discussion}
The X-ray properties of \fermi, as observed by \xmm, are  consistent overall with the identification of this object as a flaring radio-weak BL Lac \cite[]{bruni2018}.

The source exhibits a strong X-ray variability during the \xmm\ exposure, both in flux ($F_{\textrm{var}} \simeq 49$ \%\  in the 2--10 keV band) and spectral shape, on timescales of $\sim 10$ ks. The X-ray spectrum is clearly harder when the source is brighter, a behaviour typical of high-energy peaked BL Lacs \cite[]{pian2002,brinkmann2003,kraw2004,ravasio2004,gliozzi2006,zhang2006}. 
In the synchrotron or inverse Compton scenario, this variability behaviour is generally understood in terms of a competition between acceleration and cooling processes \citep[e.g.][]{tashiro1995,kirk1998}:
since higher energy electrons cool faster, the spectrum steepens when the injection rate decreases.
The mean X-ray luminosity $\nu L_{\nu}|_{1\textrm{keV}}$ is \expo{1}{45} \lumcgs, a typical value for BL Lacs \citep[e.g.][]{donato2001}.
We note that the \xmm\ observation corresponds to an intermediate flux state of the source as recorded by the \swift/XRT monitoring, which will be discussed in a forthcoming work. The latest XRT exposures (up to July 2018) show that the source was still active, albeit in a lower flux state (Fig. \ref{fig:lc_xrt}).

The X-ray spectrum is nicely described by a broken power law;
the spectral break occurs at around 2.7 keV and is consistent with being constant, regardless of the flux/spectral variations. The change in photon index is 0.6--0.7 (\ser{0.60}{0.03} on average), i.e. 
not far from
the theoretical value of 0.5 expected in the presence of a radiative cooling break, either due to synchrotron or to inverse Compton in the Thomson limit \cite[]{kardashev1962,georg2001}. 
Synchrotron cooling is expected to produce a break at an energy of
\begin{equation}
h \nu_{\textrm{br}} \simeq 1.06 \left( \frac{B}{30 \textrm{G}} \right)^{-3} \left( \frac{\tesc}{300 \textrm{s}} \right)^{-3}  \textrm{eV}
,\end{equation}
where $\tesc$ is the escape time from the acceleration zone \cite[see also][]{ponti2017}. 
Assuming a magnetic field of 0.32 G, as derived from the SED modelling in \cite{bruni2018}, we can estimate $\tesc = 2$ ks.
In the hypothesis that the acceleration is due to Bohm-type diffusion \cite[e.g.][]{stawarz2002}, the escape time is related to the size $D$ of the acceleration region as 
\begin{equation}
\tesc \sim \frac{3 D^2}{\eta r_g c}
,\end{equation}
where $\eta$ is a numerical factor and $r_g$ is the gyration radius. If $\eta=1$ (Bohm limit), 
\begin{equation}
\tesc \sim \expom{6}{16} \, \Gamma^{-1}  \left( \frac{B}{1 \textrm{G}} \right) \left( \frac{D}{\tentom{15} \textrm{cm}} \right)^{2} \textrm{s}
\end{equation}
and we obtain $D\sim \tentom{9}$ cm. This spatial scale is much smaller than that required by causal connection, namely $c \tesc \simeq \expom{6}{13}$ cm \cite[see also][]{tammi&duffy2009}. Moreover, $D$ is only \tento{-5} in units of the Schwarzschild radius. The synchrotron cooling scenario thus appears unlikely to explain the observed spectral break. The radiative cooling could instead be  dominated by Comptonization of external radiation fields, in particular from the broad-line region (BLR). When the cooling is dominated by inverse Compton of BLR photons, the spectral break can be evaluated as \cite[][]{sikora2009}
\begin{equation}
h \nu_{\textrm{br}} \simeq 4 \left( \frac{\varGamma}{20} \right)^{2} \textrm{keV}
,\end{equation}
where $\varGamma$ is the bulk Lorentz factor.
Assuming a break energy of 2.66 keV (from the fit of the average spectrum), we obtain $\varGamma=16.3$, in perfect agreement with the value estimated by \cite{bruni2018}. 
Since the break energy is consistent with being constant during the observation, we can infer that the bulk Lorentz factor is also consistent with being constant on a 50 ks timescale.

In addition to the primary continuum, a bumpy excess is found in the soft X-ray band. Intrinsic absorption is not able to reproduce this component, which  phenomenologically is nicely described by a blackbody with a temperature of 0.22 keV.
Its average 0.5--2 keV flux is \expo{5}{-12} \fluxcgs, corresponding to $\sim 20$ \%\ of the flux of the primary power law in the same band. Moreover, the
flux is found to decrease by $\sim 40$ \%\ during the softer/fainter state, although there is no clear trend with the primary flux. 
We can interpret this soft emission component as being due to bulk Comptonization of low-energy photons in the jet. This process, first proposed by \cite{begelman1987}, consists in inverse Compton scattering of optical/UV diffuse radiation by cold electrons moving with a relativistic bulk velocity \cite[see also][]{sikora1994}. As later noted by \cite{celotti2007}, a bulk Comptonization component could explain the flattening in soft X-rays seen in the spectra of some sources, like the high-redshift blazar GB~1428+4217 \cite[see also][]{worsley2004}. In general, however, this feature can be difficult to observe because it can be transient and/or masked by the primary continuum \cite[]{celotti2007} or even absent if all the jet electrons are highly relativistic \cite[]{sikora2009}. 

So far, evidence of bulk Comptonization has been found in a handful of powerful sources like 4C~04.42 \cite[]{derosa2008}, PKS 1510-089 \cite[]{kataoka2008}, and 4C~25.05 \cite[]{kammoun2018}, which belong to the class of flat-spectrum radio quasars. 
To our knowledge, \fermi\ is the first BL Lac showing possible evidence of bulk Comptonization.
This could be due to the relative weakness of the jet, which does not outshine the bulk Comptonization component (although it dominates the X-ray emission). Further X-ray observations of other radio-weak BL Lacs could thus prove valuable to search for signatures of bulk Comptonization. 
According to the analysis of \cite{celotti2007}, an excess in the soft X-rays indicates that the most likely source of soft photons is the 
BLR rather than the accretion disc. This is due to a geometrical reason, since BLR photons are mostly seen head-on, and thus they can reach the maximum blueshift, i.e. $\sim \varGamma^2$, where $\varGamma$ is the bulk Lorentz factor. 
Following \cite{celotti2007}, $\varGamma$ can be estimated by approximating the BLR emission as a blackbody spectrum. 
The relation between the temperature of the blackbody-like Comptonized emission $\ktbb$ and the BLR temperature $kT_{\textrm{BLR}}$ is thus $\ktbb \simeq \varGamma^2 kT_{\textrm{BLR}}$. Assuming $\varGamma = 16.3$, we obtain the plausible value $T_{\textrm{BLR}} \sim \tentom{4}$ K \citep[e.g.][]{peterson2006}.
In principle, the BLR is illuminated by both the disc and the jet \citep{ghisellini1996}. However, given the low luminosity of the disc estimated by \cite{bruni2018}, the dominant source of illumination is likely the jet itself. 
Finally, the BLR clouds may not be the only source of seed photons: significant contributions could come from the molecular torus and intracloud plasma \citep{celotti2007}.

In the optical/UV band, we find evidence of a significant emission component in addition to the host galaxy.
The broad-band (optical/UV to X-rays) spectrum is nicely described by a double-broken power law with a low-energy break at 5 eV;
if due to the jet, it could indicate a double-broken power law shape for the underlying electron distribution. An alternative origin involving blackbody-like disc emission is statistically disfavoured and is not consistent with the multiwavelength SED discussed by \cite{bruni2018}. However, given all the uncertainties, we cannot rule out a significant contribution from the accretion disc or the BLR. Moreover, the bulk Compton spectrum of disc photons is expected to peak in the UV band \citep{celotti2007}. 
In any case, we can compare the luminosity of the optical/UV emission with that of the soft X-ray excess to further assess the validity of the bulk Comptonization scenario. First, the luminosity of the double-broken power law in the 0.001--0.1 keV band is \expo{2.8}{45} \lumcgs. We can assume that around 10 \%\ of this luminosity is reprocessed and re-emitted by the BLR \citep{ghisellini1996}. We thus obtain a luminosity of the same order of magnitude as that of the soft X-ray excess, which is  plausible in the bulk Compton scenario \citep{celotti2007}.
The mean UV luminosity $\nu L_{\nu}$ at \tento{15} Hz is \expo{2}{44} \lumcgs\ and is in good agreement with the SED reported in \cite{bruni2018}.
A detailed analysis of the optical properties of the source is beyond the scope of this paper and is deferred to a dedicated study of all the optical follow-up data.

Finally, the soft X-ray spectrum, both time-averaged and time-resolved, displays an emission line at 0.70 keV (rest-frame) that can be tentatively attributed to \ion{Fe}{xvii}. 
This feature  could be due to thermal emission from the hot interstellar medium of the host galaxy. This is often observed in giant, non-active elliptical galaxies with temperatures generally ranging from 0.5 to 1 keV \citep{matsumoto1997,xu2002,tamura2003,werner2009,ogorzalek2017}. In the present case, we can estimate a temperature of 0.28 keV
and an observed luminosity of in the 0.1--10 keV range $L_X\suprm{gal} =\expom{5.8}{43}$ \lumcgs. This can be compared to the luminosity of the host galaxy as derived from the historical magnitude reported in the Two Micron All Sky Survey \citep[2MASS;][]{2mass}. The 2MASS $H$-band magnitude of 2MASX~J15441967-0649156 is 14.3, yielding for the luminosity in the $H$ band $\log(L_H\suprm{gal}/L_{H \odot})=11.5$ in units of solar luminosities in the same band.
Both $L_X\suprm{gal}$ and $L_H\suprm{gal}$ are within the ranges of observed values for early-type, non-active galaxies \citep{ellis2006}, although $L_X\suprm{gal}$ is relatively high. 
However, we cannot exclude that the emission line is an artefact of an imperfect modelling of Galactic absorption.
This issue might deserve further investigations through deeper X-ray observations since the detection of a soft X-ray emission line would be a quite surprising result for a BL Lac, if confirmed.\\ 
\indent
\fermi\ is clearly a very peculiar jetted source. On the one hand, the gamma-ray flare was not followed by any enhancement of the radio activity, at least on a timescale of a  few months; this could be due either to an inefficient jet collimation or to a large distance between the radio and gamma-ray emitting regions \citep{bruni2018}. 
On the other hand, a new, bright X-ray source became visible  $\sim 10$ days after the gamma-ray flare, and remained active for more than one year (May 2017--July 2018). With a 0.1--2.4 keV flux of \expo{1.7}{-11} \fluxcgs\ during the \xmm\ observation, this source is two orders of magnitude brighter than the limiting flux of the \rosat\ all-sky survey \citep[\tento{-13} \fluxcgs;][]{boller2016}. This would imply an X-ray flare of a factor larger than 100.
However, we can set a more conservative upper limit of $\sim \expom{5}{-13}$ \fluxcgs\ taking into account the \rosat\ exposure in the field of the source \citep{salvato2018}.
The observed flux of the soft excess in the same band is \expo{2}{-12} \fluxcgs, i.e. at least a factor of 4 brighter than the \rosat\ limiting flux. On the other hand, the putative thermal emission from the host galaxy has a flux of \expo{7}{-13} \fluxcgs, namely comparable with the \rosat\ limiting flux. It is thus possible that this component was present (as expected if it were due to the host galaxy), but not detected with \rosat.\\ 
\indent 
The multiwavelength properties of \fermi\ seen so far indicate that this source is a low-power, radio-weak BL Lac that underwent a major outburst leading to the ignition of strong X-ray activity. 
In its pre-flare state, \fermi\ was only a weak radio source with no other signatures of jet activity. 
This might suggest that \fermi\ could mostly stay in a quiescent state, and become active following a sporadic ejection event \citep[see also][]{gal-yam2002}.
Additional monitoring will be needed to follow the evolution of the source, which might show a progressive decline and/or new episodes of activity. This would be desirable also to better constrain the variability of the different spectral components, especially of the soft excess/bulk Comptonization feature. Further observations of this and other similar sources will be needed to determine the 
rate and power of flares
in this kind of objects and their incidence among the overall population of blazars.

\section{Summary}\label{sec:summary}
We reported on the first \xmm\ observation of the gamma-ray transient \fermi. Our main findings can be summarized as follows:
\begin{itemize}
        \item[-] The X-ray variability is strong and indicative of a harder-when-brighter behaviour, consistent with the BL Lac nature of this object;
        \item[-] The X-ray primary continuum is nicely described by a variable broken power law, likely due to synchrotron emission from the jet; the spectral break at $\sim 2.7$ keV is consistent with external Compton cooling, possibly from BLR photons;
        \item[-] The soft X-ray spectrum shows a blackbody-like soft excess, whose flux is around 20 \%\ of that of the primary power law, consistent with bulk Comptonization of BLR/intracloud plasma photons along the jet;
        \item[-] The optical/UV to X-ray spectrum is nicely described by a double-broken power law, suggesting that the jet dominates the UV emission; however, a significant contribution from the disc and/or the BLR cannot be ruled out;
        \item[-] A soft X-ray emission line is detected at 0.7 keV, which can be tentatively attributed to thermal emission of the host galaxy.
\end{itemize}
The peculiar properties of \fermi\ motivate further observations of this and other radio-weak BL Lacs, which will improve our understanding of this new class of objects and, more generally, of the physics of extragalactic jets.
\section*{Acknowledgements}
We thank the referee for a helpful and constructive review that significantly improved the manuscript.
We acknowledge financial support from ASI under contracts ASI/INAF 2013-025-R1 and I/037/12/0. This work is based on observations obtained with \textit{XMM-Newton}, an ESA science mission with instruments and contributions directly funded by ESA Member States and NASA.
\bibliographystyle{aa}
\bibliography{mybib.bib}
\appendix
\section{RGS spectrum}\label{sec:rgs}
\begin{figure} 
        \includegraphics[width=\columnwidth]{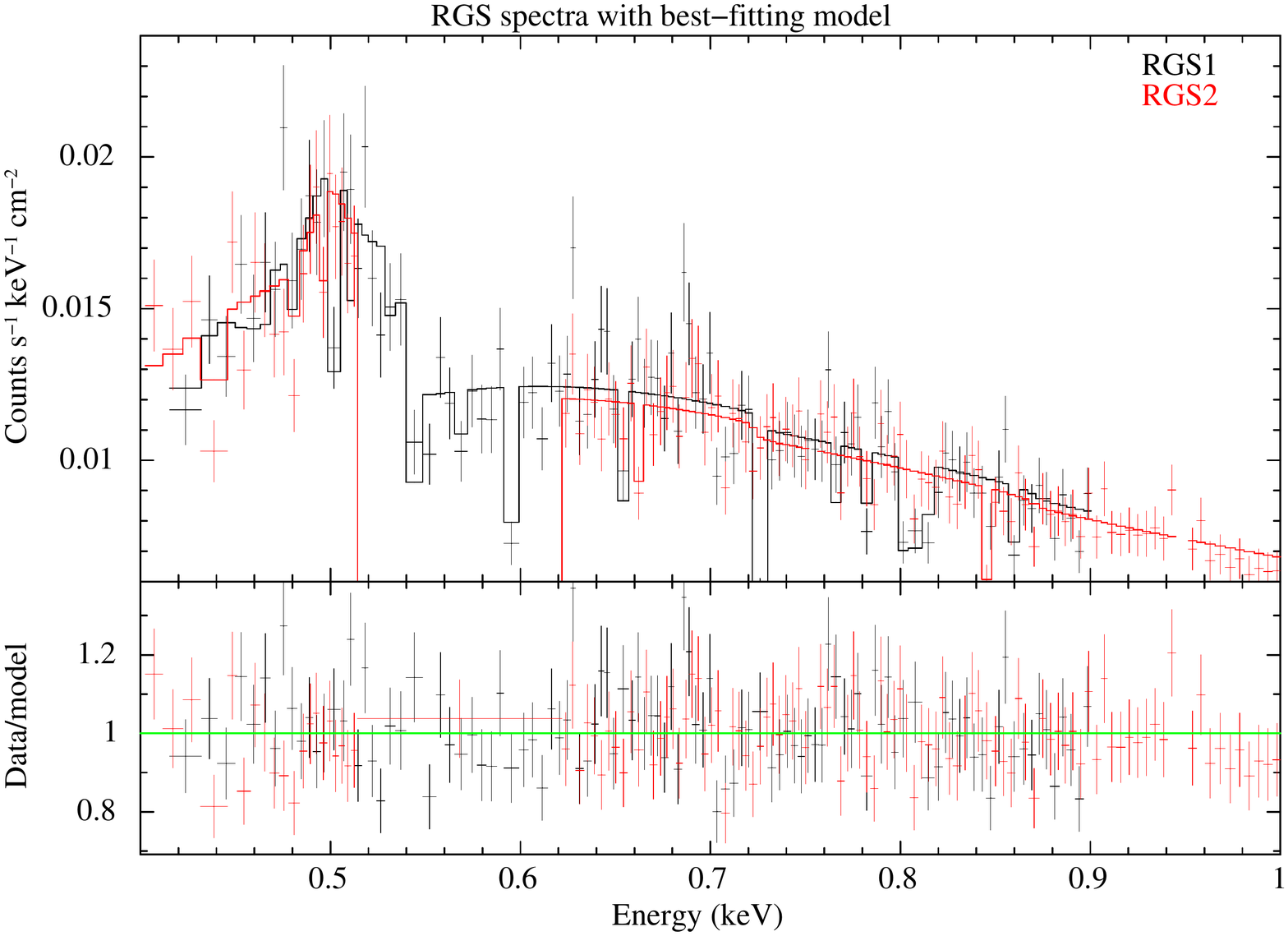}
        \caption{Upper panel: Time-averaged RGS spectrum (0.4--1 keV) with best-fitting model. Lower panel: Data-to-model ratio. The data were binned for plotting purposes. \label{fig:rgs}}
\end{figure}

We analysed the data from the Reflection Grating Spectrometer \cite[RGS;][]{RGS} with the main goal of further constraining the putative emission line at 0.6 keV seen in the pn spectrum. The RGS data were extracted using the standard \textsc{sas} task \textsc{rgsproc}. 
To obtain a good signal-to-noise ratio, we fitted the time-averaged spectra from the two detectors RGS1 and RGS2 in the 0.3--2 keV band. 
The spectra were not binned and were analysed using the $C$-statistic \cite[][]{cstat}.
Starting with a simple power law, we found a quite poor fit with $\rcash=5870/5255$. Prompted by the results of the pn data analysis of Sect. \ref{sec:pn_average}, we included a blackbody, finding an improved fit ($\rcash=5648/5253$, i.e. $\dcash/\ddof=-222/-2$). 
We obtained a further improvement by adding a Gaussian emission line, with a free energy and intrinsic width ($\rcash=5633/5250$, i.e. $\dcash/\ddof=-15/-3$). The RGS spectra with best-fitting model are plotted in Fig. \ref{fig:rgs}. The line energy was found to be \ser{0.585}{0.007} keV, the intrinsic width \aer{11}{9}{5} eV, and the flux \aer{1.4}{0.9}{0.7} in units of \tento{-4} photons \sqcm\ \pers. As such, the line has a flux consistent with that measured with the pn, while its energy is smaller by about 0.1 keV. 

This mismatch could be due to cross-calibration issues between the two instruments \citep{kirsch2004}. The line energy measured by RGS is close to, although not fully consistent with, that of the \ovii\ K$\alpha$ triplet \citep[0.561--0.574 keV;][]{atomdb}. 
This makes the identification of the atomic transition responsible for the emission line uncertain. 
We performed  a similar analysis of the MOS time-averaged spectra in the 0.3--2 keV band, and we note that the line is found to have an energy of \aer{0.71}{0.01}{0.02} keV, i.e. consistent with that found with the pn.
Finally, as in Sect. \ref{sec:pn_average}, we switched to the elemental abundances of \cite{asplund2009}. We obtained a slightly worse fit with $\rcash=5640/5250$ ($\dcash=+7$). Including a further \textsc{phabs} component, we obtained $\rcash=5634/5249$ and a column density of \serexp{4}{2}{20} \sqcm. The parameters of the Gaussian line are not significantly altered, thus indicating that the line detection in RGS is not affected by the assumptions on elemental abundances. 

\section{Testing the log-parabola}
As an alternative to the BPL model discussed in Sect. \ref{sec:analysis}, we fitted the pn data with a log-parabolic spectral law (LP) with $\nu F_{\nu}$ normalization (\eplogpar\ in \xspec; \citealt{massaro2004}, \citealt{tramacere2007}). In this model, the spectral shape is determined by the curvature parameter $\beta$ and by $E_p$, the peak energy in $\nu F_{\nu}$. 

\begin{figure} 
        \includegraphics[width=\columnwidth]{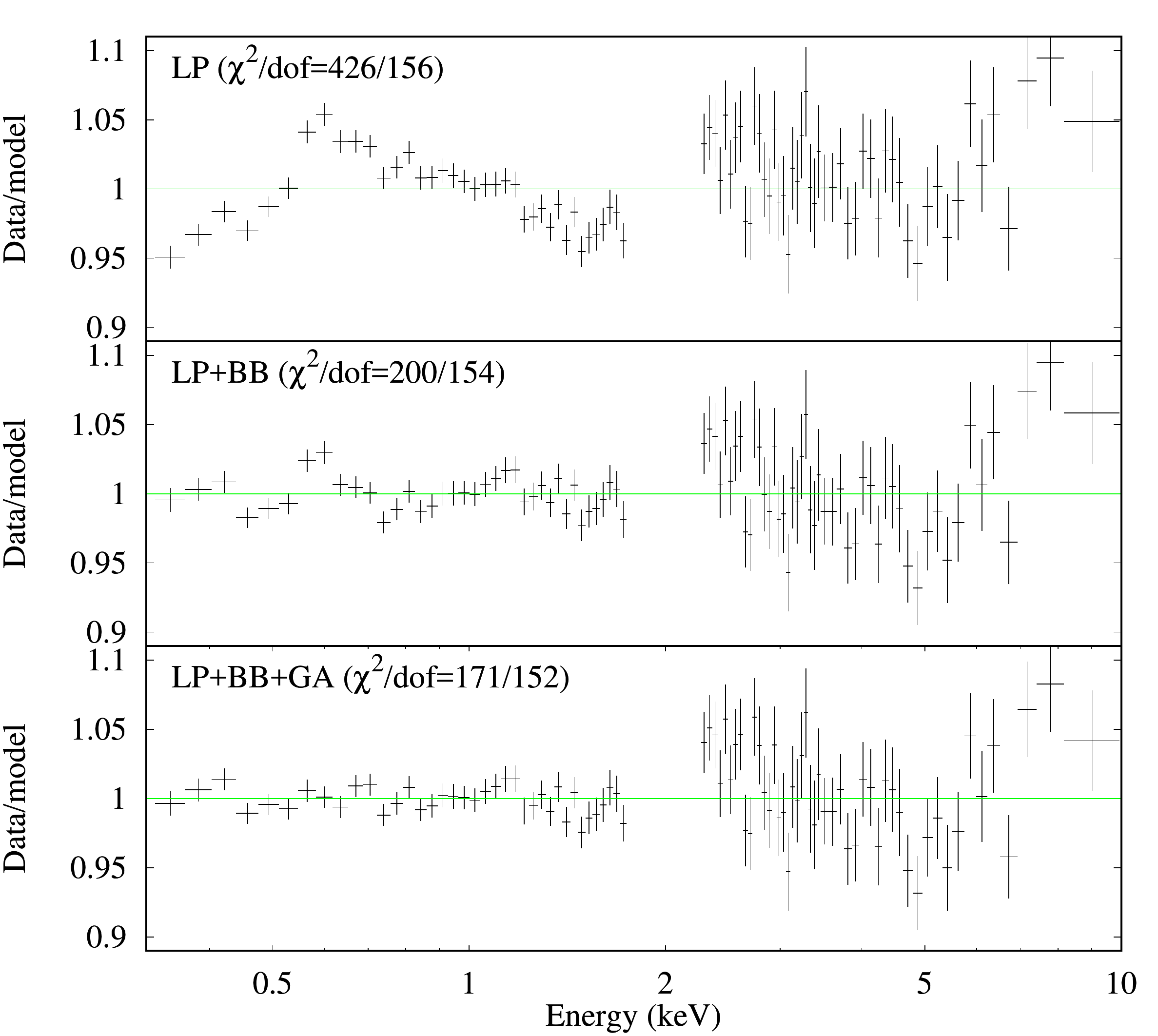}
        \caption{Residuals of the fit of the time-averaged pn spectrum in the 0.3--10 keV band with the LP model (Sect. \ref{sec:pn_average}). Top panel: Simple log-parabola. Middle panel: Log-parabola plus blackbody. Bottom panel: Log-parabola plus blackbody and a Gaussian emission line at 0.7 keV. The data were binned for plotting purposes. \label{fig:ratio_average_2}}
\end{figure}

We started by fitting the time-averaged pn spectrum, as in Sect. \ref{sec:pn_average}. Fitting with only \eplogpar, we found an unacceptable fit with $\rchisq=426/156$. Including a blackbody, as in the BPL fit, we found a significant improvement ($\rchisq=200/154$, i.e. $\dchi/\ddof=-226/-2$). Finally, including a narrow Gaussian line at 0.7 keV we found a further improvement ($\rchisq=171/152$, i.e. $\dchi/\ddof=-29/-2$). However, the final LP fit is statistically worse than the BPL best fit, especially in the harder band ($\dchi/\ddof=+29/+1$). The residuals for the different LP fits are shown in Fig. \ref{fig:ratio_average_2}.

Then, we tested the LP model on the OM+pn data, as in Sect. \ref{sec:pn_average}.
The LP model alone is not adequate to reproduce the broad-band spectrum, as even after re-fitting it yields
$\rchisq=629/151$.  We obtained a strong improvement by adding a \diskbb\ component;
however, the fit is still relatively poor ($\rchisq=167/149$), mainly because of significant residuals above 2 keV (as in Fig. \ref{fig:ratio_average_2}). 

\begin{figure} 
        \includegraphics[width=\columnwidth]{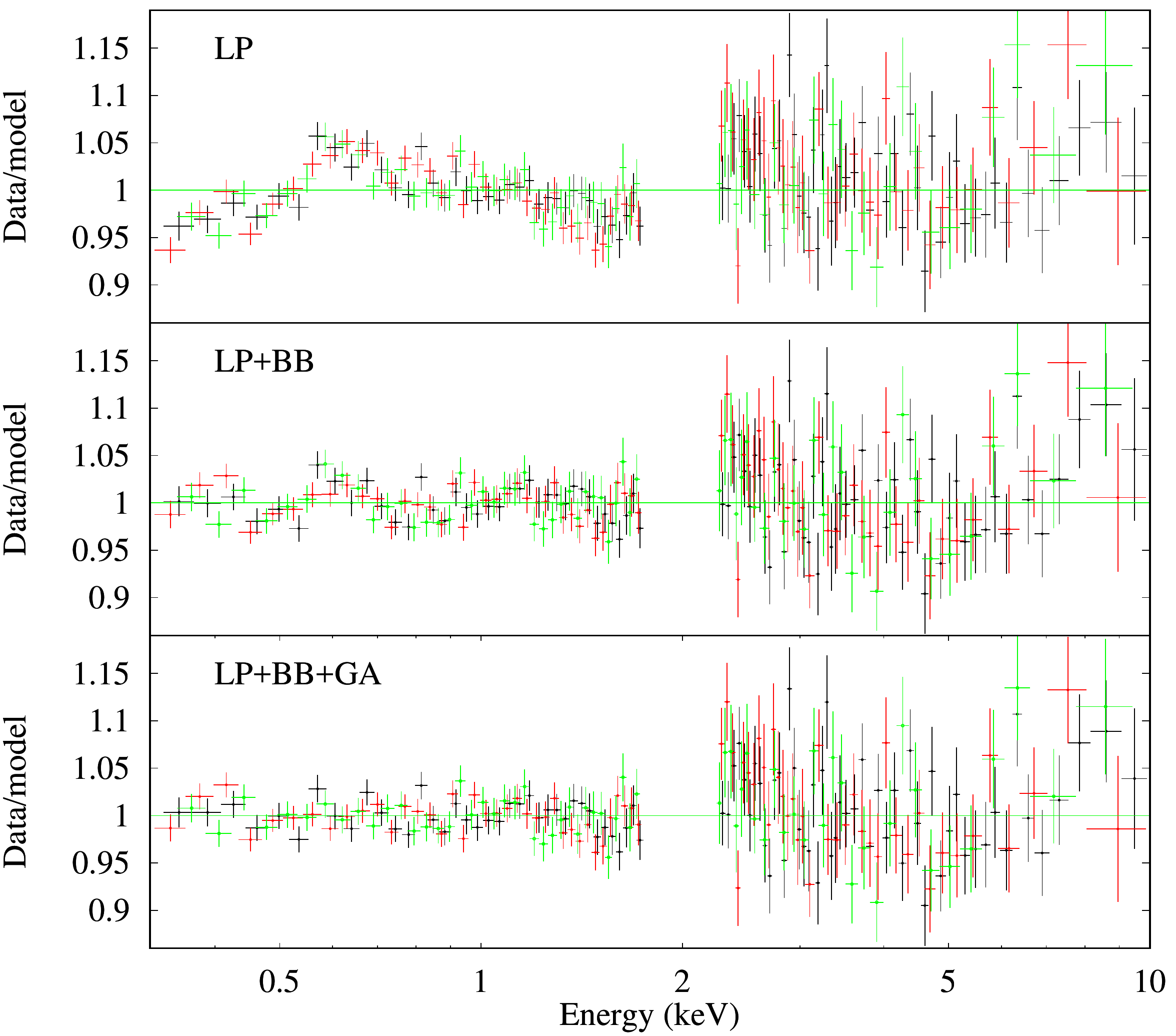}
        \caption{Residuals of the fits of the three pn spectra in the 0.3--10 keV band with the LP model. Upper panel: Simple log-parabola. Middle panel: Log-parabola plus blackbody. Lower panel: Log-parabola plus blackbody and Gaussian line. The data were binned for plotting purposes. \label{fig:res_ep}}
\end{figure}

Finally, we performed a time-resolved spectral analysis of the three intervals A, B, and C, as  in Sect. \ref{sec:time-res}. The curvature parameter of the LP and its peak energy, as well as the normalization, were left free to vary among the different intervals.
Fitting with only \eplogpar, we found an unacceptable fit with $\rchisq=702/425$. Including a blackbody, we found a significant improvement ($\rchisq=495/421$, i.e. $\dchi/\ddof=-207/-4$). Finally, including a narrow Gaussian line at 0.7 keV we found a further improvement ($\rchisq=467/419$, i.e. $\dchi/\ddof=-28/-2$). Still, the fit is statistically worse than the BPL best fit ($\dchi/\ddof=+26/+3$). The residuals for the different models are shown in Fig. \ref{fig:res_ep}, while the best-fitting parameters are listed in Table \ref{tab:fits2}.

\begin{table}
        \begin{center}
                \caption{ \label{tab:fits2} Best-fitting parameters for the time-resolved pn spectra (LP model: \textsc{eplogpar+bbody+zgauss}). (t) denotes a parameter tied among the different intervals.}
                \begin{tabular}{ l c c c } 
                        \hline  
                        & A & B & C \\
                        $E_p$ (keV)&\ser{2.12}{0.12}&\ser{1.17}{0.10}&\ser{0.75}{0.08} \\
                        $\beta$ &\ser{0.43}{0.04}&\ser{0.50}{0.04}&\ser{0.48}{0.03} \\
                        $N_{\textrm{LP}}$ (\tento{-3}) &\ser{14.0}{0.1}&\ser{10.4}{0.2}&\ser{8.0}{0.3} \\
                        \noalign{\medskip}
                        $\ktbb$ (keV)&\ser{0.175}{0.008}&(t)&(t) \\     
                        $N_{\textrm{BB}}$ (\tento{-5}) &\ser{4.8}{1.4}&\ser{5.3}{1.2}&\ser{2.3}{0.8} \\      
                        \noalign{\medskip}
                        $E_{\textrm{GA}}$ (keV)&\ser{0.72}{0.02}&(t)&(t) \\      
                        $N_{\textrm{GA}}$ (\tento{-4}) &\ser{1.1}{0.4}&(t)&(t) \\                              
                        \noalign{\medskip}
                        $\rchisq$&467/419&&\\                                                                   
                        \hline          
                \end{tabular}
        \end{center}
\end{table} 

\end{document}